# OPTIMAL SCHEDULING OF FILE TRANSFERS WITH DIVISIBLE SIZES ON MULTIPLE DISJOINT PATHS

*Mugurel Ionut Andreica*

Computer Science Department, Politehnica University of Bucharest
Splaiul Independentei 313, 060042, Bucharest, Romania
phone: + (40) 722803022, email: mugurel.andreica@cs.pub.ro
web: https://mail.cs.pub.ro/~mugurel.andreica

## ABSTRACT

In this paper I investigate several offline and online data transfer scheduling problems and propose efficient algorithms and techniques for addressing them. In the offline case, I present a novel, heuristic, algorithm for scheduling files with divisible sizes on multiple disjoint paths, in order to maximize the total profit (the problem is equivalent to the multiple knapsack problem with divisible item sizes). I then consider a cost optimization problem for transferring a sequence of identical files, subject to time constraints imposed by the data transfer providers. For the online case I propose an algorithmic framework based on the block partitioning method, which can speed up the process of resource allocation and reservation.

## 1. INTRODUCTION

The importance of data transfer scheduling techniques in achieving good communication performance has increased recently, with the world-wide development and deployment of distributed systems, services and applications. In this paper I study several offline and online data transfer scheduling problems and propose novel, efficient techniques for addressing these problems. First, I present an efficient heuristic algorithm for scheduling files with divisible sizes on multiple disjoint paths, in order to maximize the total profit. This problem is equivalent to the multiple knapsack problem with divisible item sizes. Then, I present an optimal algorithm for minimizing costs when a sequence of identical files must be transferred from a source to a destination, subject to time constraints imposed by the data transfer providers. I also propose an online algorithmic framework for the block partitioning method, which can be used to efficiently handle online resource allocation and reservation requests.

The rest of this paper is organized as follows: in Sections 2 and 3 I discuss the offline scheduling problems I mentioned above and present the developed solutions. In Section 4 I propose an algorithmic framework for online resource allocation and reservation. In Section 5 I discuss related work and in Section 6 I draw some conclusions.

## 2. MAXIMUM PROFIT DATA TRANSFERS

We are given n file transfer requests. For each request i, its file size ($sz_i>0$) and profit ($p_i>0$) are known. Each file must be transferred between the same source and destination. We consider the file sizes sorted in ascending order $sz_1 \leq sz_2 \leq \ldots \leq sz_n$. The file sizes are integers and divisible, i.e. $sz_i = q_i \cdot sz_{i-1}$ ($2 \leq i \leq n$), where $q_i \geq 1$ is an integer number. Each file transfer must be scheduled non-preemptively on one of the k paths available. The paths are disjoint and identical, except that each path j is available only during a time interval $[0, T_j]$. All the paths have unit transfer rate, so the time taken to transfer a file with size $sz_i$ is $sz_i$ time units. A file transfer request may be accepted or rejected. Accepting a request i means assigning it a path j and a time interval $[t, t+sz_i]$ fully included in $[0, T_j]$. At any moment, at most one file can be transferred on a path, i.e. the time intervals of the requests assigned to the same path must be disjoint. The total profit is the sum of the profits brought by each accepted request (if a request is rejected, it contributes nothing to the total profit). Obviously, we would like to accept those requests which bring a maximum total profit. This problem is equivalent to the multiple knapsack problem with divisible item sizes. Each path j is a knapsack of a given capacity $T_j$. The file transfer requests are items whose sizes are divisible and we are interested in finding a maximum profit subset of items, such that each item in the set is placed in some knapsack and the sum of the item sizes in any knapsack does not exceed its capacity. The multiple knapsack problem is NP-hard, thus a polynomial time algorithm is unlikely to exist. Even for this particular case with divisible item sizes, we present only a pseudopolynomial $O(n \cdot S \cdot min\{n, S \cdot log(S)\})$ time algorithm, where S is the maximum size of an item. A direct solution obtained by extending the standard dynamic programming algorithm for the single knapsack case takes $O(n \cdot max\{T_j\}^k)$ time (where k is the number of knapsacks) and computes a multidimensional array $P_m[i, s_1, s_2, \ldots, s_k]$=the maximum profit which can be achieved by choosing a subset of the first i items and filling each knapsack j up to size $s_j$ (at most). We have $P_m[0, s_1, \ldots, s_k]=0$ (for all the values $s_j$) and

$$P_m[i, s_1, \ldots, s_k] = \max \begin{cases} P_m[i-1, s_1, \ldots, s_k] \\ p_i + \max \begin{cases} P_m[i-1, s_1 - sz_i, s_2, \ldots, s_k], \\ P_m[i-1, s_1, s_2 - sz_i, \ldots, s_k], \\ \ldots, \\ P_m[i-1, s_1, s_2, \ldots, s_k - sz_i], \end{cases} \end{cases}$$

For $P_m[i, s_1, \ldots, s_k]$, the choices are to either ignore the $i^{th}$ item or place it in one of the k knapsacks (the item can be placed in knapsack j if $s_j \geq sz_i$). The maximum profit is given by $P_m[n,T_1,\ldots,T_k]$. However, this solution is inefficient. Fater algorithms make use of heuristics. The most natural heuristic is the following one, based on a greedy algorithm:

**Greedy1MultipleKnapsack(item_set, knapsack_set):**
k=|knapsack_set|
*fill the **first knapsack** optimally with a subset **item_sol** of the items*
**if** (*k=1*) **then**
  **return** *profit(item_sol)*
**else if** (|*item_set \ item_sol*|>0) **then**
  **return** *profit(item_sol) + Greedy1MultipleKnapsack(item_set \ item_sol, knapsack_set \ {first knapsack})*

Other heuristic algorithms consist of sorting the items according to some criterion (e.g. profit/size) and inserting them using the *First Fit* heuristic. I will now present a very different approach, which provides the optimal solution in many cases. We will split the items into groups: two items belong to the same group if they have the same size; thus, all the items in group i have size $sg_i$. We consider the groups sorted in decreasing order of the item sizes, i.e. $sg_1 > sg_2 > \ldots > sg_G$ (where G is the total number of distinct item sizes). Within a group i, the items are sorted in decreasing order of their profits, i.e. $pr_{i,1} \geq pr_{i,2} \geq \ldots \geq pr_{i,ni}$, where $n_i$ is the number of items in group i and $pr_{i,j}$ is the profit of the $j^{th}$ item in the $i^{th}$ group. In the first step of the algorithm, we will insert the items into the knapsacks using the *First Fit* heuristic. The items are traversed in increasing order of the group number and, within a group, in increasing order of the item number. For each item (i,j) (the $j^{th}$ item in the $i^{th}$ group), if it can be inserted into a knapsack p without exceeding its capacity, we will insert it into p. The knapsack index p is not important. Because the item sizes are divisible, we will be able to insert the same set of items during this first stage, no matter which knapsack p we choose for a specific item. We will then successively improve the initial solution, by replacing items with subsets of items which could not be inserted during the first stage and whose total profit is larger than the individual profit of the replaced item. The algorithm is sketched below:

**MultipleKnapsackWithDivisibleItemSizes():**
**for** *i=1* **to** *G* **do**
  **for** *j=1* **to** $n_i$ **do**
    *knapsack[(i,j)]=0*
    **for** *p=1* **to** *k* **do**
      **if** ($T_p \geq sg_i$) **then** // insert item (i,j) into knapsack p
        *knapsack[(i,j)]=p*; $T_p = T_p - sg_i$; **break**
*improved_solution=true*
**while** (*improved_solution*) **do**
  *smax=the maximum size of an item inside a knapsack*
  *nitems=0*
  **for** *i=G* **downto** *1* **do**
    **if** ($sg_i < smax$) **then**
      *nchosen=0; j=***firstItem**(*i*)
      **while** ((***isValidItem***(*i, j*)) **and** (*nchosen<floor(smax/$sg_i$)*)) **do**
        *nitems=nitems+1; cand[nitems]=(i,j)*
        *csz[nitems]=$sg_i$ ; nchosen=nchosen+1*
        *j=***nextItem**(*i, j*)
  $P_{max}[i,C]=0$, *for* $0 \leq i \leq nitems$, $0 \leq C \leq smax$
  **for** *i=1* **to** *nitems* **do**
    $P_{max}[i,C] = P_{max}[i-1,C]$, *for any* $0 \leq C \leq smax$
    **for** *C=csz[i]* **to** *smax* **do**
      $P_{max}[i,C] = \max\{P_{max}[i-1,C], P_{max}[i-1,C-csz[i]] + pr_{cand[i]}\}$
  *maxdif=max*$\{P_{max}[nitems,sg_i]-pr_{i,j} \mid knapsack[(i,j)]>0\}$
  **if** (*maxdif>0*) **then**
    $(i_r,j_r)$=*the item to be replaced (for which maxdif is maximum)*
    *Q=the subset of items in cand, corresponding to* $P_{max}[nitems,sg_{ir}]$
    **for** *(i,j)* **in** *Q* **do** *knapsack[(i,j)]= knapsack[($i_r,j_r$)]*
    *knapsack[($i_r,j_r$)]=-1; improved_solution=true*
  **else** *improved_solution=false*

At the end, for each item (i,j) we have three options:
- knapsack[(i,j)]>0, indicating the knapsack into which the item is placed
- knapsack[(i,j)]=-1 : the item was inserted inside a knapsack during the first stage, but was replaced afterwards
- knapsack[(i,j)]=0 : the item was never inserted inside any knapsack

During the second stage of the algorithm, we choose *nitems* items which have never been inserted into any knapsack and compute the maximum profit obtained by choosing a subset of these items whose sum is *sum* (for each sum=*1* to *smax*); these values are stored in $P_{max}[nitems, sum]$. We then replace an item $(i_r,j_r)$ from a knapsack for which the profit increase $P_{max}[nitems, sg_{ir}] - pr_{ir,jr}$ is maximum. The replaced item is ignored from now on, as it cannot be part of an optimal solution. By maintaining a linked list with the items in each group, from which we remove (in O(1) time) an item when it is inserted into a knapsack, we can implement the *firstItem*, *nextItem* and *isValidItem* functions in O(1) time. The optimality of the algorithm is justified by the following facts: any valid solution for the multiple knapsack can be successively improved to an optimal solution by replacing a subset of items $S_1$ in one of the knapsacks with a subset of items $S_2$ outside of any knapsack. Because the item sizes are divisible, the set $S_1$ can always contain only one item. The first stage of the algorithm takes O(n·k) time and O(n) items can be inserted then. The *while* loop can be executed a number of times equal to the number of items inserted in the first stage. Each iteration of the while loop takes $O(nitems \cdot smax)$ time. Two upper limits for *nitems* are O(n) and

$$\sum_{i=1}^{smax-1} \frac{smax}{i} = O(smax \cdot \log(smax)) \ .$$

Since *smax* is bounded by *S*, the largest size of an item, the overall time complexity is $O(n \cdot S \cdot \min\{n, S \cdot \log(S)\})$.

I compared the proposed algorithm with three other algorithms: the single knapsack extension to multiple knapsacks, the Greedy1MultipleKnapsack algorithm and a greedy algorithm which sorted the items according to several criteria and then used the First Fit heuristic. I considered many test scenarios and most of them were solved optimally by the new algorithm. However, I was also able to find test cases where the algorithm could not find the optimal solution. However, in terms of performance (quality of the obtained solution and running time), the algorithm I proposed is a clear winner, followed by the Greedy1MultipleKnapsack algorithm.

## 3. MINIMUM COST DATA TRANSFERS

We are given a sequence of n similar files, which need to be sent consecutively from a source to a destination. The transfer of each file takes 1 time unit (thus, file i is transferred from time i-1 to time i). There are k data transfer providers; a provider j charges a fixed price $C_j$ per time unit for transferring data and leases his services for at most $T_{max,i}$ time units. Because of several factors, each provider j asks that the leased time interval includes a specified time interval $[T_{1,j}, T_{2,j}]$ ($T_{2,i} - T_{1,i} \leq T_{max,i}$). Since files cannot be transferred simultaneously, the time intervals rented from each provider will be disjoint. We may also use a default network link for transferring a file i, which would cost us $L_i$. Of course, we are interested in paying the minimum total cost for the file transfers. We present here an $O(k \cdot n)$ dynamic programming algorithm for solving this problem. We will sort the data transfer providers in increasing order of $T_{2,i}$, i.e. $T_{2,1} \leq T_{2,2} \leq \ldots \leq T_{2,k}$. We will compute the values $Cmin[i,j]$=the minimum total cost for sending the first j files using a subset of the first i providers (in the sorted order). Initially, $Cmin[0,0]=0$ and $Cmin[0,j]=+\infty$, for $j>0$. For $i>0$, we have:

$$Cmin[i, j] = \min \begin{cases} Cmin[i-1, j] \\ Cmin[i, j-1] + L_j, \text{ if } (j > 0) \\ +\infty, \quad \text{if } (j > T_{1,i} + T_{max,i}) \text{ or } (j < T_{2,i}) \\ (j - T_{1,i}) \cdot C_i + \\ \min_{j-T_{max,i} \leq p \leq T_{1,i}} \{Cmin[i-1, p] + (T_{1,i} - p) \cdot C_i\} \end{cases}$$

When computing $Cmin[i,j]$, we have the choice of using the services of the $i^{th}$ data transfer provider or not. If we do not use them, then the cost is equal to $\min\{Cmin[i-1,j], Cmin[i,j-1]+L_j\}$. If we want to use the $i^{th}$ provider, but j violates the time constraints imposed by the provider $((j>T_{1,i}+T_{max,i})$ or $(j<T_{2,i}))$, then the cost is $+\infty$; otherwise, j is the end time moment of the leased time interval and we need to choose the first time moment of the interval (p). Using the equation above, an $O(k \cdot n^2)$ algorithm can be implemented easily (taking $O(n)$ time for each pair (i,j)). We will show how to compute all the values $Cmin[i,j]$ in $O(n)$ time for each value of i (thus, in $O(1)$ time for every pair (i,j)). For each $1 \leq i \leq k$, we are only interested in the values of j within the interval $[T_{2,i}, T_{1,i}+T_{max,i}]$ (the others are easy to handle); thus, we will compute an array $minp_i$, where

$$minp_i[q] = \min_{q \leq p \leq T_{1,i}} \{Cmin[i-1, p] + (T_{1,i} - p) \cdot C_i\}$$

We have $minp_i[T_{1,i}]=Cmin[i-1,T_{1,i}]$. Each of the other values can be computed in $O(1)$ time (in order, from $T_{1,i}-1$ downto $T_{2,i}-T_{max,i}$):

$$minp_i[q] = \min\{minp_i[q+1], Cmin[i-1,q] + (T_{1,i}-q) \cdot C_i\}$$

After computing the array $minp_i$ in $O(n)$ time, we can compute in $O(1)$ time each value $Cmin[i,j]$, with j in $[T_{2,i}, T_{1,i}+T_{max,i}]$: $Cmin[i,j]=\min\{Cmin[i-1,j], Cmin[i,j-1]+L_j, (j-T_{1,i}) \cdot C_i + minp_i[j-T_{max,i}]\}$. The total cost is $Cmin[k,n]$.

## 4. ONLINE RESOURCE MANAGEMENT

We consider the following scenario: a resource manager receives resource allocation and reservation requests (data transfer requests) which need to be processed in real time (as soon as they arrive or in batches). A request asks for a certain amount of resources (e.g. bandwidth), subject to several types of time constraints (e.g. fixed duration, earliest start time, latest finish time). Many models and algorithms have been developed for online scheduling problems [1]. We consider here the following assumptions: time is divided into discrete, equally-sized time slots and the resource manager must handle many requests simultaneously, providing low response times. Because of the stringent time constraints, the scheduler needs some efficient data structures to help it check if the request's constraints can be satisfied and to choose appropriate reservation parameters (if the request is accepted). In order to speed up the processing of requests, we introduce an algorithmic framework for the block partitioning method: We have an array of n cells, where each cell has a value $v_i$ (each cell corresponds to a time slot). We will divide the n cells into n/k blocks of size k (we assume that k is a divisor of n; if it is not, n can be extended to be a multiple of k or the last block may contain fewer cells). The blocks are numbered from 0 to (n/k)-1. The cells 0, …, k-1 belong to block 0, the cells k, …, 2·k-1 belong to block 1, …, the cells (i-1)·k, …, (i·k)-1 belong to block i-1. Thus, cell j belongs to block (j div k) (integer division). For simplicity, we store for each block B the first and last cells of the block (*left[B]* and *right[B]*). Using this partitioning, we can support several update and query functions in $O(k+n/k)$ time. By choosing $k=\sqrt{n}$, we have $O(k+n/k)=O(\sqrt{n})$. Queries consist of computing a function on the values of a range of cells [a,b] (range query) or on retrieving the value of a single cell (point query).

**Range Query(a, b):** compute $qFunc(v_a, v_{a+1}, …, v_b)$.

Analogously, we have point and range updates:

**Range Update(u, a, b):** $v_i=uFunc(u, v_i)$, $a \leq i \leq b$.

The *qFunc* function must be binary and associative, i.e. $qFunc(v_a,…,v_b)=qFunc(v_a,qFunc(v_{a+1},…,qFunc(v_{b-1}, v_b)..))$ and $qFunc(a,qFunc(b,c))=qFunc(qFunc(a,b),c)$. We must also have $uFunc(x,y)=uFunc(y,x)$. Only values $v_i$ with $O(1)$ size are considered (numbers and tuples with a fixed number of elements). *uFunc* and *qFunc* must be able to handle *uninitialized* arguments. If one of their arguments is *uninitialized*, they must simply return the other argument; this part will be intentionally left out of the functions' descriptions. The algorithmic framework consists of the functions from Table 1.

Table 1. Algorithmic Framework Functions

| Update Functions | Query Functions |
| --- | --- |
| BPpointUpdate | BPpointQuery |
| BPrangeUpdate | BPrangeQuery |
| BPrangeUpdatePoints | BPrangeQueryPoints |
| BPrangeUpdatePartialBlock | BPrangeQueryPartialBlock |
| BPrangeUpdateFullBlock | BPrangeQueryFullBlock |

In order to perform a range update, we will call the *BPrangeUpdate* function with the corresponding parameters (the update value u and the update interval [a,b]). This function splits the update interval into three zones: the first block $B_a$ intersected by the interval (containing the cell a), the last block $B_b$ intersected by the interval (containing the cell b) and all the blocks in between $B_a$ and $B_b$ (the inner blocks). The blocks $B_a$ and $B_b$ may not be fully contained inside the interval: they will be updated in $O(k)$ time (partial update). All the inner blocks are fully contained inside [a,b]: they will be updated in $O(1)$ time each (full update). Since there are $O(n/k)$ such blocks, the overall complexity of a range update is $O(k+n/k)$. The range query function (*BPrangeQuery*) works similarly. For each block B we will maintain two values: *uagg* and *qagg*. *uagg* is the aggregate of the update parameters of the function calls which updated all the elements of B (for which B was an inner block). uagg is reset to an *uninitialized* value on each partial update of the block. *qagg* is the answer to the query function called on all the elements of B. The point update and query functions are: *BPpointUpdate* and *BPpointQuery*. The framework also uses a "multiplication" operator *mop*, which computes the effects of an update operation upon the query result on a range of cells. This operator must exist when range queries and range updates are used together, but can be ignored otherwise. When the data structure is initialized, the *uagg* value of each block is set to *uninitialized* (*qagg* is initialized with the query result on the range of the block's cells). This framework is similar to the segment tree framework introduced in [6] and can support all the combinations of point and range query and update functions mentioned there.

**BPpointUpdate(u, i):**
$v_i = uFunc(u, v_i)$
B=the block to which the cell i belongs
qagg[B]=**BPrangeQueryPoints**(left[B], right[B])

**BPrangeUpdate(u, a, b):**
$B_a$, $B_b$=the blocks of cells a and b
**if** ($B_a = B_b$) **then**
  **if** (($a$=left[$B_a$]) **and** ($b$=right[$B_a$])) **then**
   **BPrangeUpdateFullBlock**($B_a$, u)
  **else BPrangeUpdatePartialBlock**($B_a$, u, a, b)
**else**
  **BPrangeUpdatePartialBlock**($B_a$, u, a, right[$B_a$])
  **BPrangeUpdatePartialBlock**($B_b$, u, left[$B_b$], b)
  **for** block=$B_a$+1 **to** $B_b$-1 **do**
   **BPrangeUpdateFullBlock**(block, u)

**BPrangeUpdatePoints(u, a, b):**
**for** p=a **to** b **do** $v_p$=uFunc(u, $v_p$)

**BPrangeUpdatePartialBlock(B, u, a, b):**
**BPrangeUpdatePoints**(uagg[B], left[B], right[B])
uagg[B]=uninitialized
**BPrangeUpdatePoints**(u, a, b)
qagg[B]=**BPrangeQueryPoints**(left[B], right[B])

**BPrangeUpdateFullBlock(B, u):**
uagg[B]=**uFunc**(u, uagg[B])
qagg[B]=**uFunc**(**mop**(u, 1eft[B], right[B]), qagg[B])

**BPpointQuery(i):**
B=the block to which the cell i belongs
**return uFunc**(uagg[B], $v_i$)

**BPrangeQuery(a, b):**
$B_a$, $B_b$=the blocks of cells a and b
**if** ($B_a = B_b$) **then**
  **return BPrangeQueryPartialBlock**($B_a$, a, b)
**else**
  $q_a$=**BPrangeQueryPartialBlock**($B_a$, a, right[$B_a$])
  $q_b$=**BPrangeQueryPartialBlock**($B_b$, left[$B_b$], b)
  q=uninitialized
  **for** block=$B_a$+1 **to** $B_b$-1 **do**
   q=**qFunc**(q, **BPrangeQueryFullBlock**(block))
  **return qFunc**($q_a$, **qFunc**(q, $q_b$))

**BPrangeQueryPoints(a, b):**
q=uninitialized
**for** p=a **to** b **do** q=**qFunc**(q, $v_p$)
**return** q

**BPrangeQueryPartialBlock(B, a, b):**
**BPrangeUpdatePoints**(uagg[B], left[B], right[B])
uagg[B]=uninitialized
**return BPrangeQueryPoints**(a, b)

**BPrangeQueryFullBlock(B):**
**return** qagg[B]

In the case of point queries with range updates, only the *uagg* values are meaningful; similarly, only the *qagg* values are meaningful in the case of point updates with range queries. Common update and query functions can be easily integrated into the framework. For example, with *uFunc(x,y)=(x+y)*, *qFunc(x,y)=(x+y)* and *mop(u,a,b)= u·(b-a+1)*, we can support point and range sum queries, together with point and range addition updates. For *uFunc(x,y)=x+y*, *qFunc(x,y)=min(x,y)* and *mop(u,a,b)=u*, we can support point and range minimum (or maximum) queries, together with point and range addition updates. We can also consider point and range multiplication updates, *uFunc(x,y)=x·y*, with point and range queries: *qFunc(x,y)=x·y* (with *mop(u,a,b)=$u^{b-a+1}$*), *qFunc(x,y)=min(x,y)* and *qFunc(x,y)= (x+y)* (with *mop(u,a,b)=u*). With *mop(u,a,b)=u*, we can support range queries and updates for some bit functions (where $v_i$=0 or 1). For *uFunc(x,y)=(x or y)* and *uFunc(x,y)=(x and y)*, we can have *qFunc(x,y)=(x and y)* and *qFunc(x,y)=(x or y)*. For the *and* update, we can also have *qFunc(x,y)=(x xor y)*. We can support range xor updates and queries (*uFunc(x,y) = qFunc(x,y)=(x xor y)*), but with *mop(u,a,b)=(if (((b-a+1) mod 2)=0) then 0 else u)*. In order to obtain any combination of bit functions, we notice that the result of a query depends only on the number of 0 and 1 values ($cnt_0$, $cnt_1$) in the query range: if ($cnt_1$>0) then *or* returns 1; if ($cnt_1$ mod 2=1) then *xor* returns 1; if ($cnt_0$=0) then *and* returns 1. Thus, we will work with ($cnt_0$, $cnt_1$) tuples as values. We will also consider the *conceptual values* $cv_i$, which are the numerical values we conceptually work with. We have $v_i$=(1-$cv_i$, $cv_i$). A query asks for the number of 0 and 1 conceptual values in the query range and an update changes this number according to the bit function used. Any combination of point and range queries and updates is supported with the functions below:

**bitTupleQuery(($cnt_{0,x}$, $cnt_{1,x}$), ($cnt_{0,y}$, $cnt_{1,y}$)):**
**return** ($cnt_{0,x}$+$cnt_{0,y}$, $cnt_{1,x}$+$cnt_{1,y}$)

**bitTupleUpdate((1-u, u), ($cnt_0$, $cnt_1$), func):**
**if** *(func=and)* **and** *(u=0)* **then return** *($cnt_0$+$cnt_1$, 0)*
**else if** *(func=or)* **and** *(u=1)* **then return** *(0, $cnt_0$+$cnt_1$)*
**else if** *(func=xor)* **and** *(u=1)* **then return** *($cnt_1$, $cnt_0$)*
**else return** *($cnt_0$, $cnt_1$)*

If the update function has the effect of setting all the values in a range to the same value *s* (range set), we will again need to work with tuples: the values $v_i$ and the update parameters u will have the form *(numerical value, time_stamp)*. We need to have a *timestamp()* function which returns increasing values upon successive calls. We can use a global counter as a time stamp, which is incremented at every call. The initial numerical values are assigned an initial time stamp and every update parameter gets a more recent time stamp. The update function is:

**uFunc(($w_x$, $t_x$), ($w_y$, $t_y$)):**
**if** ($t_x > t_y$) **then return** *($w_x$, $t_x$)* **else return** *($w_y$, $t_y$)*

With these definitions, a point query function call on a position i will return the last update parameter of an interval containing that position.

A useful range query function (used together with point updates) is finding the maximum sum segment (interval of consecutive cells) fully contained in a range of cells [a,b] (see [9] for this problem without updates). Conceptually, the value of a cell i is a number $cv_i$, but in the framework we will use tuples consisting of 4 values: *(totalsum, maxlsum, maxrsum, maxsum)*. Assuming that these values correspond to an interval of cells [c,d], we have the following definitions:

$$totalsum = \sum_{p=c}^{d} cv_p \qquad maxlsum = \max_{c-1 \leq q \leq d} \sum_{p=c}^{q} cv_p$$

$$maxrsum = \max_{c \leq q \leq d+1} \sum_{p=q}^{d} cv_p \qquad maxsum = \max_{\substack{c \leq q \leq d \\ q-1 \leq r \leq d}} \sum_{p=q}^{r} cv_p$$

In the framework, a value $v_i$ will be a tuple corresponding to the interval [i,i]. If $cv_i$<0, then $v_i$=($cv_i$, 0, 0, 0); otherwise, $v_i$=($cv_i$, $cv_i$, $cv_i$, $cv_i$). The point update function changes the value of $cv_i$ of a cell i and then recomputes $v_i$. The *qFunc* function is given below:

**qFunc(($t_x$,$ml_x$,$mr_x$,$m_x$), ($t_y$,$ml_y$,$mr_y$,$m_y$)):**
 **return** ($t_x$+$t_y$, max{$ml_x$, $t_x$+$ml_y$}, max{$mr_y$, $t_y$+$mr_x$}, max{$m_x$, $m_y$, $mr_x$+$ml_y$})

We can use the range set update together with the range maximum sum segment query – this combination is not supported by the framework in [6]. Conceptually, each cell has a numerical value $cv_i$. Practically, the framework's values $v_i$ will be tuples of the following form *(totalsum, maxlsum, maxrsum, maxsum, time_stamp)*. The update, query and multiplication functions are given below. We must notice that the fundamental combination (range set update, range sum query) is also solved. However, I could not find suitable function definitions for the combination (range addition update, range maximum sum segment query).

**uFunc((total$_x$, ml$_x$, mr$_x$, m$_x$, t$_x$), (total$_y$, ml$_y$, mr$_y$, m$_y$, t$_y$)):**
**if** ($t_x$>$t_y$) **then return** *(total$_x$, ml$_x$, mr$_x$, m$_x$, t$_x$)*
**else return** *(total$_y$, ml$_y$, mr$_y$, m$_y$, t$_y$)*

**qFunc((total$_x$, ml$_x$, mr$_x$, m$_x$, t$_x$), (total$_y$, ml$_y$, mr$_y$, m$_y$, t$_y$)):**
**return** *(total$_x$+total$_y$, max{ml$_x$, total$_x$+ml$_y$}, max{mr$_y$, total$_y$+mr$_x$}, max{m$_x$, m$_y$, mr$_x$+ml$_y$}, max{t$_x$, t$_y$})*

**mop((total$_x$, ml$_x$, mr$_x$, m$_x$, t$_x$), a, b):**
**return** *((b-a+1)·total$_x$, (b-a+1)·ml$_x$, (b-a+1)·mr$_x$, (b-a+1)·m$_x$, t$_x$)*

The framework's behaviour can be improved by adding a dirty flag to each block. With the dirty flag, the *qagg* value will be recomputed only "on demand" and not after every point or partial block update. We only need to replace the functions *BPpointUpdate*, *BPrangeUpdatePartialBlock* and *BPrangeQueryFullBlock* with the following definitions:

**BPpointUpdate(u, i):**
$v_i$=uFunc(u,$v_i$)
*B=the block to which the cell i belongs*
*dirty[B]=true*

**BPrangeUpdatePartialBlock(B, u, a, b):**
**BPrangeUpdatePoints**(*u, a, b*)
*dirty[B]=true*

**BPrangeQueryFullBlock(B):**
**if** (dirty[B]) **then**
 **BPrangeUpdatePoints**(*uagg[B], left[B], right[B]*)
 *uagg[B]=uninitialized*
 *qagg[B]=***BPrangeQueryPoints**(*left[B], right[B]*)
 *dirty[B]=false*
**return** *qagg[B]*

## 5. RELATED WORK

Optimal high multiplicity scheduling algorithms for file transfers with divisible sizes, with the objective of minimizing the makespan, were presented in [2]. Related bin packing, knapsack and multiple knapsack problems were studied in [3,4,5]. Although the single knapsack problem with divisible item sizes was solved in [5], the corresponding multiple knapsack version does not seem to have been addressed so far. The algorithmic framework for the block partitioning technique is based on a similar framework for the segment tree data structure, presented in [6]. The block partitioning technique has been used in order to enhance the performance of range queries and updates in many domains, particularly in dynamic OLAP data cubes [7,8].

## 6. CONCLUSIONS

In this paper I presented two efficient algorithms for two offline data transfer scheduling problems. The first one is equivalent to the multiple knapsack problem with divisible item sizes, for which I am unaware of any previous results. The second one is a minimum cost optimization problem, for which the proposed dynamic programming algorithm is optimal. For the online case I proposed an algorithmic framework for the block partitioning technique. The framework allows to efficiently handle pairs of query and update operations whose usefulness is unquestionable in several classes of real-time resource managers and bandwidth brokers.